\documentclass[journal=jacsat,manuscript=article]{achemso}

\usepackage[version=3]{mhchem} 

\usepackage{xcolor}
\usepackage{gensymb}

\author{Mackenzie O’Keefe}
\affiliation{Department of Physics, University of Massachusetts Boston, Boston, MA 02125, USA}
\author{Jane Bernadette Denise M. Garcia}
\affiliation{Department of Physics and Astronomy, Johns Hopkins University, Baltimore, MD 21218, USA}
\author{Abeco J.  Rwakabuba}
\affiliation{Department of Physics, University of Massachusetts Boston, Boston, MA 02125, USA}
\author{Timothy M. Otchy}
\affiliation{Department of Biology and Neurophotonics Center, Boston University, Boston, MA 02215, USA}
\author{Daniel A. Beller}
\affiliation{Department of Physics and Astronomy, Johns Hopkins University, Baltimore, MD 21218, USA}
\email{d.a.beller@jhu.edu}
\author{Mohamed Amine Gharbi}
\affiliation{Department of Physics, University of Massachusetts Boston, Boston, MA 02125, USA}
\email{mohamed.gharbi@umb.edu}

\title[An \textsf{achemso} demo]
  {Templated self-assembly of gold nanoparticles in smectic liquid crystals confined at 3D printed curved surfaces\footnote{Mackenzie O’Keefe and Jane Bernadette Denise M. Garcia contributed equally to this work.}}


\begin{document}







\begin{abstract}
  The fabrication of assembled structures of topological defects in liquid crystals (LCs) has attracted much attention during the last decade, stemming from  the potential application of these defects in modern technologies. A range of techniques can be employed to create large areas of engineered defects in LCs, including  mechanical shearing, chemical surface treatment, external fields, or geometric confinement. The technology of 3D printing has recently emerged as a powerful method to fabricate novel patterning topographies inaccessible by other microfabrication techniques, especially  confining geometries with curved topographies. In this work, we show the advantages of using 3D-printed curved surfaces and controlled anchoring properties to confine LCs and engineer new structures of topological defects, whose structure we elucidate by comparison with a novel application of Landau-de Gennes free energy minimization to the smectic A-nematic phase transition. We also demonstrate the ability of these defects to act as a scaffold for assembling gold (Au) nanoparticles (NPs) into reconfigurable 3D structures. We discuss the characteristics of this templated self-assembly (TSA) approach and explain the relationship between NP concentrations and defect structures with insights gained from numerical modeling. This work paves the way for a versatile platform of LC defect-templated assembly of a range of functional nanomaterials useful in the field of energy technology. 
\end{abstract}

\section{Introduction}
The field of nanotechnology has been rapidly growing in the last decade.  It has attracted enormous attention, particularly after the discovery of new nanomaterials and the development of diverse ways to manipulate them \cite{Yin2020}. The interest also stems from the potential of nanomaterials in producing new applications useful in a wide range of emerging fields, such as energy technology \cite{Pomerantseva2019}, bioengineering \cite{Harish2022}, sensing \cite{Gloag2019,Li2019,Chang2019}, and optics \cite{Wang2020,Fathima2021,Zhao2022}. For this reason, the scientific community has strived to develop new techniques to fabricate and manipulate nanocomposites. One class of these techniques is the top-down approach, which exploits the patterning of bulk materials to organize nanosized structures \cite{Kang2023}. Examples include the techniques of photolithography \cite{Hughes2017}, soft lithography \cite{Auzelyte2012}, scanning lithography \cite{Garno2003}, nanocontact printing \cite{Lee2010}, laser machining \cite{Yang2019},  and deposition \cite{Jiang2014}. Another set of approaches are described as bottom-up methods, in which hierarchical nanostructures are assembled by building upon single atoms and molecules \cite{Kumar2018,Andreo2022}. When these two kinds of methods are combined, they give rise to an innovative approach for nanofabrication, known as the templated-self-assembly (TSA) technique \cite{Cheng2006},  where the top-down helps the bottom-up to create specific structures \cite{Isaacoff2017}.\\   

Liquid crystals (LCs) are complex fluids that have long been of interest because of their reconfigurable properties and their fast response to external stimuli. These materials have also attracted significant attention because of their topological defects that are easy to engineer and capable of guiding the assembly of different classes of functional nanomaterials \cite{Wang2015,Gharbi2016,Honglawan2015,Lee2016,Senyuk2012,Li2017,Kim2011,Yoon2007,Coursault2012}. The use of LCs and their topological defects as a template for bottom-up designs has also opened new avenues for the development of new materials of both fundamental and technological interest. For example, they have been used as effective platforms for fabricating microlens arrays \cite{Kim2012, Serra2015}, templates for soft lithography \cite{Kim2010}, and matrices for nanoparticle (NP) assembly \cite{Do2020,Jeridi2022}. \\

In this work, we propose a new mechanism of TSA, for which we combine bottom-up self-assembly with top-down patterned templates to create reconfigurable structures of LC defects capable of manipulating and organizing nanomaterials. In particular, we use the smectic LC phase, a class of soft materials characterized by a lamellar structure consisting of rod-shaped molecules arranged in parallel layers. Different classes of smectics are distinguished by the orientation of these molecules relative to the layer normal direction. In this study, we focus on the smectic A (SmA) phase that is characterized by a perpendicular orientation of molecules within the layers and often exhibits distinctive defect structures called focal conic domains (FCDs). In an FCD, the smectic layers wrap around two defect curves, an ellipse and a hyperbola each passing through a focus of the other, that contain singularities of layer curvature. Close-packed arrays of FCDs self-assemble when a SmA LC is confined with \textit{hybrid anchoring} conditions, in which a perpendicular orientation is imposed at one interface through \textit{homeotropic} anchoring while tangential orientations are imposed at the opposite interface through \textit{degenerate planar} anchoring \cite{Yoon2007,Kim2010}.  Surface topography strongly influences the placement of FCDs \cite{kim2009confined, honglawan2011pillar,honglawan2013topographically}. When one of the confining interfaces has a curved geometry, the FCDs assemble into 3D hierarchical structures that can be manipulated by changing the surface geometry's curvature, periodicity, and/or anchoring conditions \cite{Beller2014,Gharbi2015,Preusse2020}.  \\ 

Curvature is a fundamental concept useful in many fields, such as chemistry, condensed matter, and soft matter physics. This is because it introduces specific limitations that can impact the assembly of molecules by restricting the interaction patterns between them. Additionally, curvature can influence the self-assembly behavior of different types of nanomaterials by affecting their packing geometry and stability. Other examples include the impact of curvature on lipid bilayer kinetics \cite{Parthasarathy2006}, the phase behavior of 2D nanosystems \cite{Guillemeney2022}, the properties of self-assembled monolayers \cite{Browne2011}, the dynamics of Brownian colloids \cite{Liu2021}, the interaction of anisotropic particles at fluid interfaces \cite{Cavallaro2011}, and the organization of topological defects in LCs, both passive \cite{Lopez2011} and active \cite{Ellis2017}. This is also true for biological systems such as red blood cells, for which curvature plays a critical role in the rapid gas exchange between hemoglobin and the surrounding medium to enable flexible migration into various vessels \cite{SilvaM2010}. The concept of curvature is also essential for understanding the behavior of materials with layered structures. A good example is the case of block copolymers, usually used in patterning applications or the engineering of nanostructures \cite{Vega2013,ThiVu2018}. These systems show different curved geometries depending on the volume fraction of each block \cite{Xiang2005}. SmA LCs offer another important example, as their bulk elasticity causes their layers to curve in precisely predictable ways, which can be selected through appropriate boundary conditions  \cite{Seok2017}.  \\

The control of curvature fields along with anchoring conditions could lead to powerful mechanisms to engineer director fields in LCs and assemble their defects. With the recent developments of microfabrication techniques and the emergence of the 3D nanoprinting technology, it is now possible to create curved geometries with complex shapes that were not achievable with traditional microfabrication methods \cite{Pearre2019}. This opens the doors for new ways to confine LCs and harness the effects of curvature on the behavior of soft materials with lamellar order. In this study, we extend this approach to controllably create defects in LCs by confining a SmA film under hybrid anchoring at a 3D-printed surface with double undulations (\textit{i.e.}\ height oscillations in the $X$ and $Y$ directions). We thus use curvature fields to engineer the assembly of FCDs.  We then demonstrate how these defect structures can be utilized as a scaffold to manipulate the assembly of gold (Au) NPs. We also discuss the features of this approach by analyzing the effect of Au NP concentration on the properties of the smectic film and the organization of defects.  \\   

\section{Results and discussion}
\subsection{Confinement of the LC mixture at 3D printed curved surfaces}
\begin{figure}
 \centering
  \includegraphics[scale=0.5]{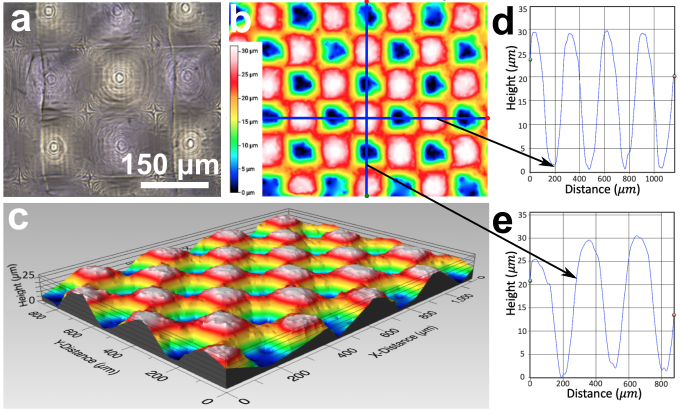}
  \caption{Characterization of the curved PDMS surface used to confine the LC mixture. (a) Optical image of the PDMS film. (b) Interferogram of the PDMS surface. (c) 3D reconstruction of the PDMS surface. Profile of the surface in the $X$ (d) and $Y$ (e) directions. The double undulations have peak-to-peak amplitudes of $A$=25$\pm$2$\mu$m and a wavelength of $\lambda$=300$\pm$2$\mu$m.  }
  \label{Fig1}
\end{figure}

We confine a 4-pentyl-4-cyanobiphenyl (5CB)/ 4'-octyl-4-cyanobiphenyl (8CB) mixture at a ratio of 20/80 by weight percent (wt.$\%$), between a PDMS film that presents undulations in both $X$ and $Y$ directions and a coverslip treated with polyvinyl alcohol (PVA).  Figure~\ref{Fig1} shows the optical image and the 3D profile of the PDMS film used to confine the mixture. The PDMS film was replicated from a 3D printed master mold that has undulations with peak-to-peak amplitudes of $A$=25$\pm$2$\mu$m and a wavelength of $\lambda$=300$\pm$2$\mu$m.  The anchoring of the 5CB/8CB mixture is perpendicular at the PDMS surface and degenerate planar (tangential) at the coverslip,  resulting in a hybrid alignment of the LC film.  This alignment is confirmed by analyzing the texture and the defect structure of the mixture in the flat region in the same sample, where the LC film is uniform in both the N and smectic phases.\\

The 5CB/8CB system presents an isotropic phase at temperatures above $T_{NI}=39.3\pm0.2~\degree$C, a nematic (N) phase between $T_{NI}$ and $T_{SN}=16.4\pm0.4~\degree$C, a smectic A (SmA) phase between $T_{SN}$ and $T_{CS}=5.4\pm0.2~\degree$C, and a crystal phase below $T_{CS}$. Figure~\ref{Fig2} shows optical images of the mixture near the N-SmA phase transition. \\

We used a mixture of 5CB and 8CB instead of pure 8CB because the latter presents a strong viscosity at room temperature, which could be a disadvantage with Au NPs. We found that adding a small amount of 5CB to the 8CB preserves the N and SmA mesophases, but modifies their phase transition temperatures, leading to a N phase at room temperature and a SmA phase at lower ones \cite{ternet1999flow}. We also noticed that the phase diagram of this system is very sensitive to the addition of 5CB. We tested different ratios and found the optimal concentration to be 20/80 wt.$\%$, because higher 5CB concentrations suppress the SmA phase. \\

\subsection{Characterization of the defect structure in the LC}

We begin our characterization by analyzing the texture of the LC mixture near the N-SmA phase transition and comparing the patterns of LC defects at flat and double-undulated surfaces. The goal of this analysis is to understand the effect of curvature on the structure and assembly of topological defects in the N and SmA phases. When cooling the mixture from the isotropic to the N phase, we observe the appearance of disclinations lines that tend to shrink over time if the LC is maintained in its N phase. The energetic cost of these defect lines gives them an effective line tension \cite{degennes_prost1995}, so the LC tends to shrink and eliminate them over time in order to minimize its free energy. However, this is not the case in the double undulated confinement. The defects could remain for a significant amount of time, suggesting that curvature is responsible for creating and stabilizing defects in N LCs. \\ 

\begin{figure}
  \includegraphics[width=\linewidth]{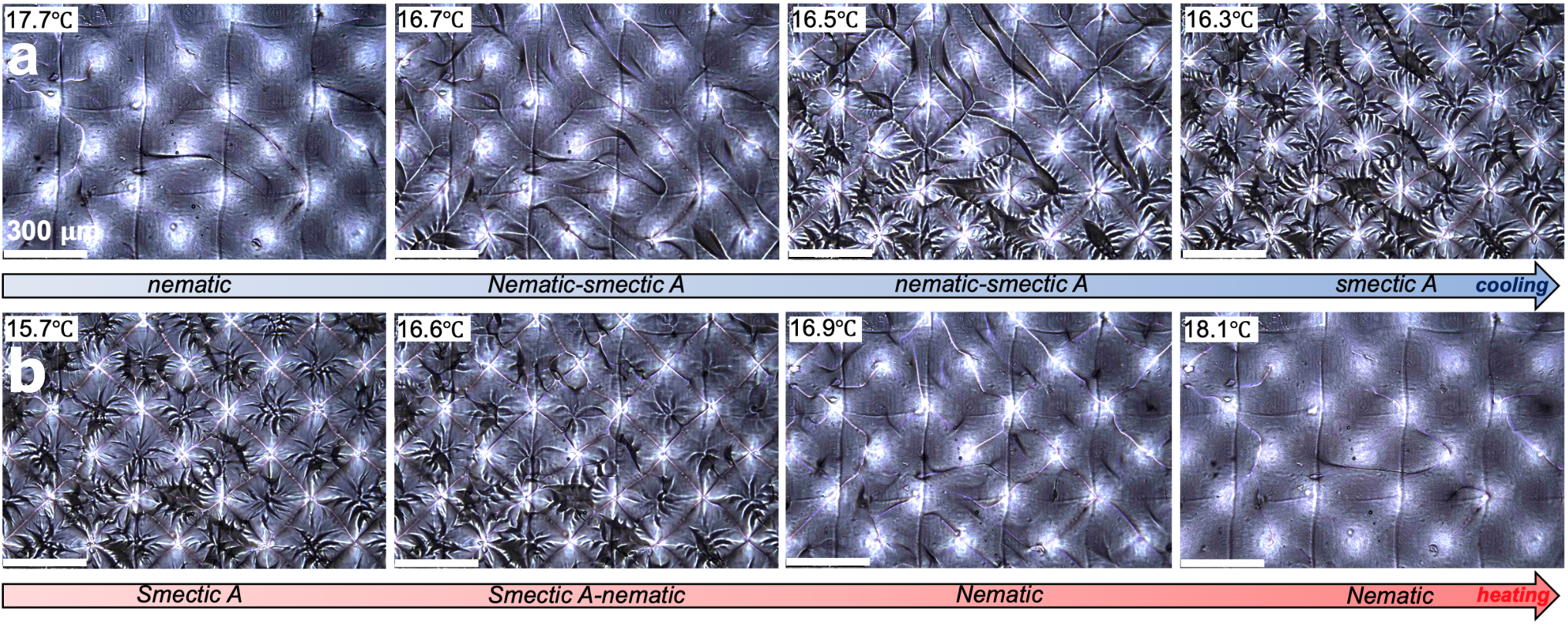}
  \caption{Behavior of the LC mixture near the N-SmA phase transition. Optical images showing the textures of the LC mixture during cooling from the N to the SmA phase (a), and during heating from the SmA to the N (b). The images also show how disclinations and FCDs are formed and arranged during the phase transition. We note here that the lines that appear both vertically and horizontally in all figures are artifacts of the 3D printing process.} 
  \label{Fig2}
\end{figure}

As the system is cooled toward the SmA phase, more disclinations appear at precise locations forming periodic patterns. Additionally, FCDs of different sizes emerge, forming hierarchical assemblies of larger and smaller defects around these disclinations. This process is reversible: disclinations in the N phase re-emerge at approximately the same locations after the system is heated and cooled through  many cycles, as shown in Figure \ref{Fig2}, exhibiting a form of geometric memory across the phase transition \cite{Suh2019}. \\

\begin{figure}
\centering
  \includegraphics[width=0.8\linewidth]{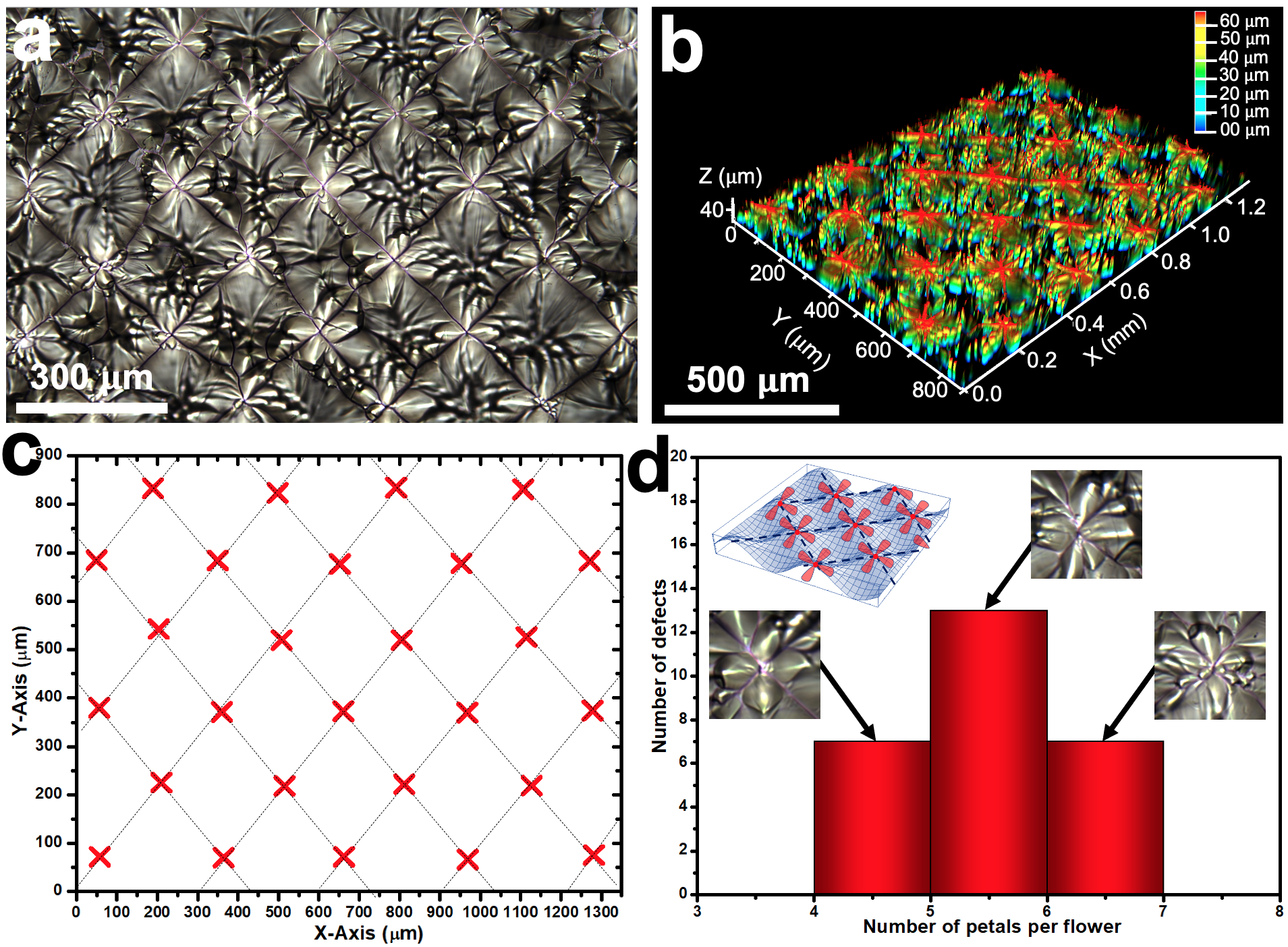}
  \caption{Characterization of the defect structure of the SmA LC confined at a 3D printed double undulated surface. (a) Optical image of the smectic patterns showing the formation of parallel and perpendicular defect lines that appear to intersect each other near the peaks of the undulations. The image also shows the formations of flower patterns of FCDs around the disclinations. (b) 3D reconstruction of the defect assembly in the SmA obtained by scanning the sample along the $Z$ axis. (c) Measure of the smectic flower center positions. The dashed lines are virtual lines that indicate the structure of disclinations connecting the positions of the flower centers. (d) Histogram of the number of petals per smectic flower.  Inset images show examples of flowers with four, five, and six petals. The inset sketch indicates the position of these flowers with respect to the double undulated PDMS surface.     }
   \label{Fig3}
\end{figure}

To better understand the formation of defects in the SmA phase and the role of the 3D printed surface, we characterized the structure of the defects as a function of the confining surface profile.  Figure ~\ref{Fig3}-a  shows a typical optical image of the SmA confined at a double-undulated surface. From above, each FCD appears as an ellipse of high eccentricity, distinguishing these defects as elliptic-hyperbolic FCDs as opposed to the zero-eccentricity toric FCDs \cite{kleman_lavrentovich_2003}.  Figure~\ref{Fig3}-b is the corresponding 3D reconstruction of the smectic texture obtained by scanning the sample along the $z$-axis. These measurements indicate that the disclinations form a crosshatch grid of lines that appear to intersect near the coverslip, above the crests in the double-undulated surface.  
We are unable to determine the $z$-separation of the defects, which means we cannot confirm if they are really intersecting or not. The FCDs assemble in groups resembling petals of a flower, centered at these disclination intersection sites. These results confirm the correlation between the morphology of the confining surface and the assembly of defects.

Additionally, we measured the number of petals present in the flower patterns and found that they all have between 4 and 6 petals, as shown in the histogram of Figure~\ref{Fig3}-d. However, the majority of the flowers present 5 petals. We believe that the variability in petal numbers is linked to the thickness of the smectic film between the crest of the double undulation and the coverslip's surface, as FCDs' size and number are sensitive to this parameter. Since the thickness of our samples varies slightly, which is challenging to regulate experimentally, the number of petals may differ from one region to another, as seen in Figure~\ref{Fig3}-a, from the bottom right to the top left. These results suggest that further control over FCD assembly structures may be obtainable by systematically varying parameters that we do not investigate here, including sample thickness and the amplitude and wavelength of the double undulations.   \\

\begin{figure}
\centering
  \includegraphics[scale=0.33]{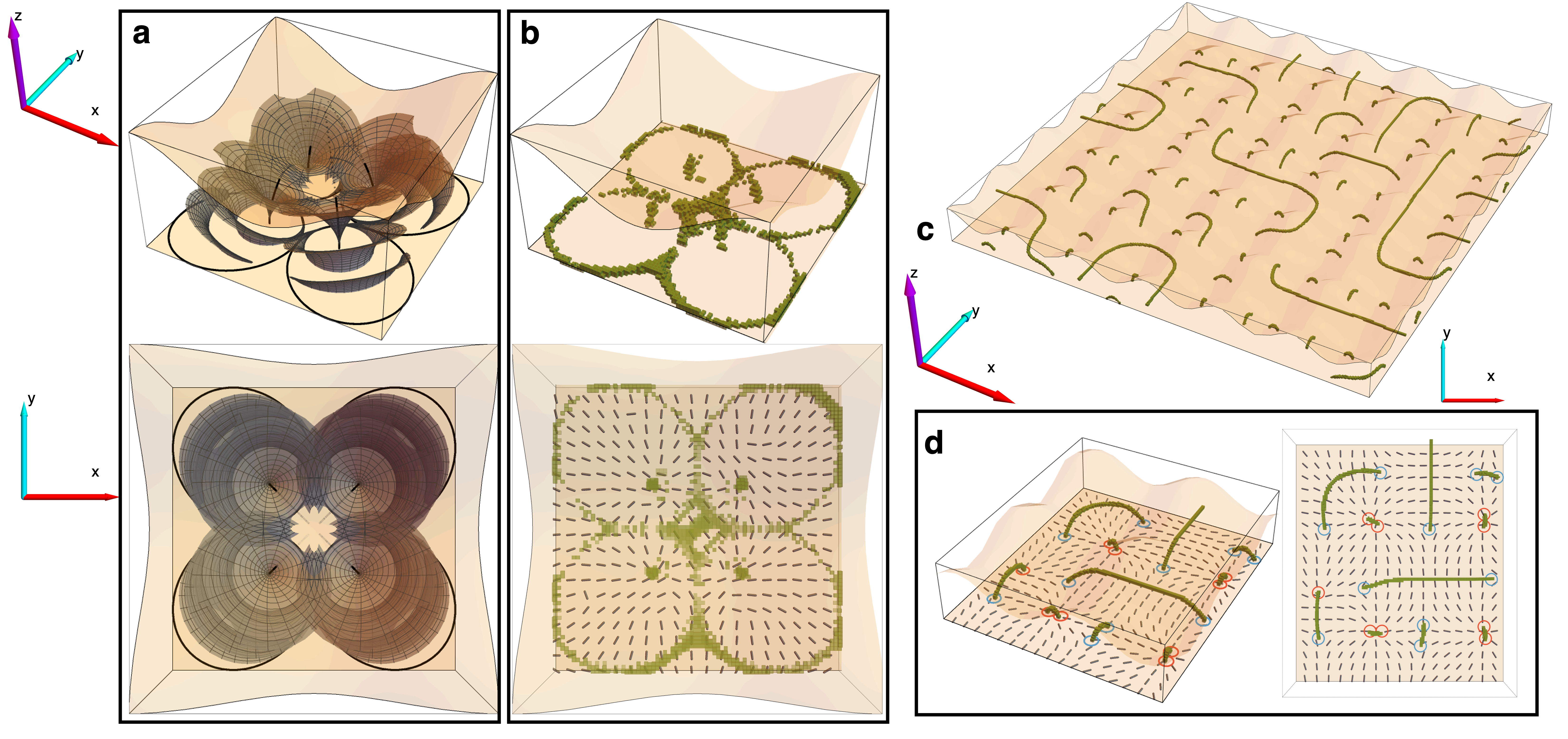}
  \caption{Numerical modeling of defect evolution in double-undulated confinement at the SmA-to-N phase transition. (a) Schematic of the positions of the FCDs used in the first application of the LdG free energy minimization. (b) Defects (green)  arising after the first LdG relaxation, in which the director is held fixed inside the FCDs. Gray rods represent the director field on the flat surface with degenerate planar anchoring. (c) A metastable configuration of disclination lines after final application of LdG free energy minimization, modeling a state deep in the N phase. (d) Two zoomed-in views of a subset of the defects in boxed in (c). Defect endpoints intersecting the flat surface are encircled with blue ($-1/2$ winding) and red ($+1/2$ winding) circles. Point-like $\pm 1$ defects are visible in their split-core forms as a short disclination half-loops with winding number $\pm 1/2$ at both endpoints.}
   \label{fig:LdGDefectEvolution}
\end{figure}

In order to better understand the LC defect configurations in the double-undulated confinement, and their reversible transformations at the N-SmA phase transition, we turn to numerical modeling. We perform a multi-step application of Landau-de Gennes (LdG) free energy minimization, accounting for the FCD arrangement in the SmA phase and the changing elasticity in the N close to $T_{SN}$. Figure~\ref{fig:LdGDefectEvolution}-a shows the assumed arrangement of four elliptic-hyperbolic FCDs in each unit cell of the double-undulation, which we take as part of our model's initial condition, choosing the most symmetrical FCD petal arrangement for simplicity. With the N director field within the FCDs known analytically \cite{kleman_lavrentovich_2003}, we first relax the director field only outside of the FCDs, and with bend elastic constant larger than the splay elastic constant, $k_3 \equiv K_3/K_1 = 4$, to approximate the smectic configuration. 

Starting from this FCD director configuration, we perform a second relaxation at $k_3=4$ over the entire domain to produce a modeled N configuration just above $T_{SN}$. Finally, a third relaxation at $k_3=1.67$ evolves the system into a metastable state deep inside the N phase. We find a large number of possible metastable  disclination arrangements, with the defect lines'  endpoints lying on the flat (degenerate planar anchoring) surface. An example of such a metastable array is shown in Figures~\ref{fig:LdGDefectEvolution}-c and d. This highly multistable scenario is in stark contrast with the disclination configuration arising under single-undulated confinement, as investigated experimentally in \cite{Preusse2020} and numerically in the Supporting Information (Figure~S1), where the N LC exhibits a unique stable state with parallel, straight disclinations. 

Furthermore, the curvature of the double undulated surface geometry determines the local structure of nearby defects: Disclination line endpoints where the N director has  $+1/2$ winding occur opposite the undulation crests, where the LC thickness is minimum, while those with $-1/2$ winding lie opposite the saddle points (median thickness) of the double undulated surface. This indicates a relationship between surface defect ``charge'' at one interface and the Gaussian curvature of the nearby opposite surface, extending a connection between surface curvature and defect charge known previously from in-surface \cite{turner2010vortices,ellis2018curvature} and bulk \cite{tran2016lassoing} behaviors. 

Additionally, we observed the formation of point-like $\pm 1$ defects at the interstices of the ellipses of the FCDs. Defects with winding number $+1$ were observed in the N state at the locations of the interstitial centers of converging FCD flower groups in the smectic initial condition, while defects with $-1$ charge were observed at some of the sites of contact between FCDs belonging to different groups. These point-like defects appear as split-core short disclination lines \cite{Tasinkevych2012} and are similar to those seen in previous simulations of toric FCD packings at the SmA-to-N phase transition \cite{Suh2019}. In the present context, it is interesting to note that the relationship between winding number and  Gaussian curvature at the opposite surface follows the same trend for point-like defects as for the ends of extended disclination lines: positive charges appear opposite domes where the Gaussian curvature is positive, and negative charges appear opposite saddle points where the Gaussian curvature is negative. The presence of point defects was confirmed experimentally in our system, as shown in Figure-S2. The density of point defects is lower in experiment than in the numerical results, which may indicate that a more detailed model of the SmA-to-N phase transition will be required for quantitative comparison.

\subsection{Directed assembly of Au NPs via engineered defects in the smectic LC}
It is known that topological defects in LCs present strong trapping sites for nanomaterials \cite{Wang2015,Gharbi2016,Coursault2012,Do2020,Jeridi2022}. In this section, we explore the possibility of using defects assembled at curved geometries as a scaffold to organize NPs into reconfigurable ordered structures. We used Au NPs of nominal diameter $20\,\mu\mathrm{m}$, stabilized with citrate ligands and dispersed in the LC mixture at different concentrations. The presence of ligands at the NP surfaces helps them to disperse uniformly in the LC and prevents them from forming large aggregates. The ligands also promote degenerate planar anchoring of the LC molecules at the surfaces of the NPs. In this section, we focus on the SmA phase because of its highly organized defects. Our motivation is to investigate whether NPs could conform to the structure of LC defects.

We examined various concentrations of Au NPs, ranging from 0.01 wt.$\%$ to 0.2 wt.$\%$. We first establish the behavior of our NPs in uniformly confined SmA films with hybrid alignment, where both interfaces are planar. The goal was to verify if NPs used in this study could interact with self-assembled FCDs, as previously demonstrated with other types of NPs \cite{Yoon2007, Ok2016}. Our findings show that the NPs are attracted to the FCDs and arranged in patterns that resemble the FCDs' structure. This was confirmed by observing the sample near the N-SmA phase transition (Figure~S3). These results demonstrate the potential of smectic defects in attracting the NPs and manipulating their organization.

\begin{figure}
\centering
  \includegraphics[scale=0.2]{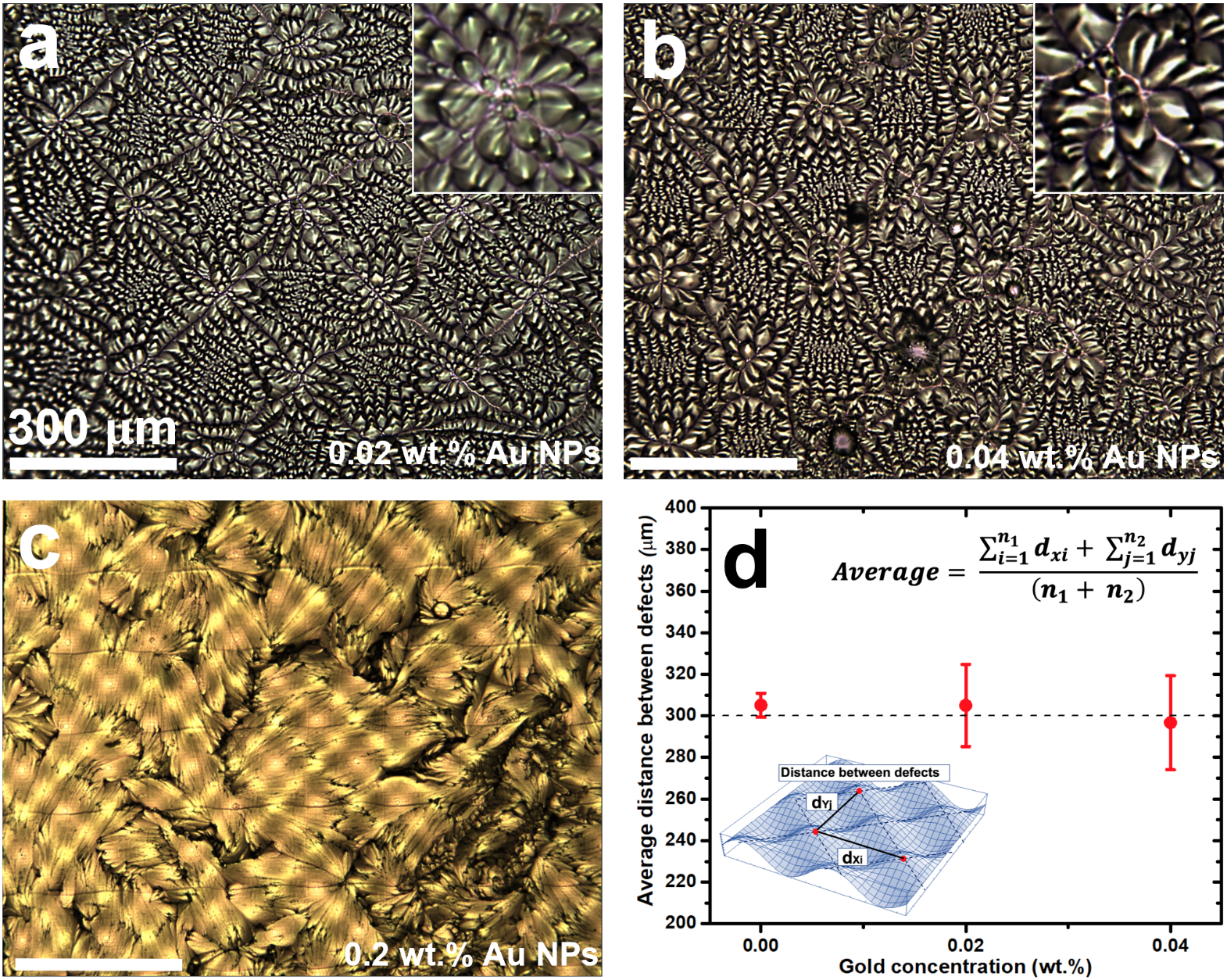}
  \caption{Texture of the SmA as a function of Au NP concentration. Defect structures of the SmA with 0.02 wt.$\%$ (a), 0.04 wt.$\%$ (b), and 0.2 wt.$\%$ Au NPs (c). When the concentration of NPs is high (c), the disclinations and FCDs disappear. This is because the NPs migrate to the boundaries and change the anchoring properties of the SmA from hybrid to planar. (d) Average distance between defect intersections as a function of Au NP concentration. The inset sketch and formula show how we measured the average distance between defect intersections. }
   \label{Fig4}
\end{figure}

Figure~\ref{Fig4} shows optical images of the  samples prepared at different NP concentrations: 0.02 wt.$\%$, 0.04 wt.$\%$, and 0.2 wt.$\%$.  Our results indicate that the presence of a low concentration of NPs doesn't affect the structure of disclinations and FCDs in the smectic film, as shown in Figure~\ref{Fig4}-a. For Au NP concentration of 0.02 wt.$\%$, the average distance between the defect lines, obtained by averaging the distances $d_X$ and $d_Y$ between the defect intersections in both $X$ and $Y$ directions (see inset of Figure~\ref{Fig4}-d), is $d\approx$ 305$\pm$20 $\mu$m, compared to $d \approx$ 305$\pm$6 $\mu$m for samples without NPs. This confirms that FCDs and disclinations created via curvature can serve as a template to guide the assembly of nanomaterials since they are strong trapping sites for NPs, as confirmed with flat samples. When the concentration of NPs is increased to 0.04 wt.$\%$, these defects continue to form ordered structures but with some deformations in their arrangements, as shown in Figure~\ref{Fig4}-b. The flower patterns of FCDs that form around the disclinations become less defined with a larger number of petals. The average distance between the disclination intersection points decreases, and its standard deviation increases ($d \approx$ 296$\pm$23 $\mu$m). This result suggests that at certain concentrations, the NPs can distort the regular defect structures of the smectic obtained without or with low NP concentrations (see Figure~\ref{Fig4}-d).   \\

When the concentration of NPs is higher than 0.2 wt.$\%$, the texture of the smectic changes drastically, as shown in Figure~\ref{Fig4}-c. The FCDs and disclinations disappear, and the effect of surface curvature apparently vanishes. The smectic no longer presents a hybrid-aligned texture, suggesting that the anchoring conditions for the LC have changed at the boundaries. This could be induced by the accumulation of a large number of NPs at the surface of PDMS and coverslip due to the strong elastic forces of the LC expelling the NPs from the bulk. The presence of NPs at the boundaries could change the orientation of smectic layers at the double undulated surface to satisfy the surface anchoring of the NPs, which is degenerate planar, rather than the homeotropic anchoring of the PDMS. This result is similar to previous work with planar anchoring fluorosilane functionalized silica (F-SiO2) NPs dispersed in semi-fluorinated smectic LC \cite{Honglawan2015}. In that case, the NPs migrate to the boundaries and form monolayers capable of changing the alignment of smectic molecules. We believe that a similar mechanism is responsible for the change in smectic alignment in our system at high NP concentration. These results show that the TSA approach we propose in this study to organize NPs into reconfigurable 3D structures is inverted at high NP concentration: Rather than SmA LC defects directing the assembly of NPs, sufficiently concentrated NPs drastically alter the assembly of the SmA LC layers. \\

\begin{figure}
\centering
  \includegraphics[scale=0.5]{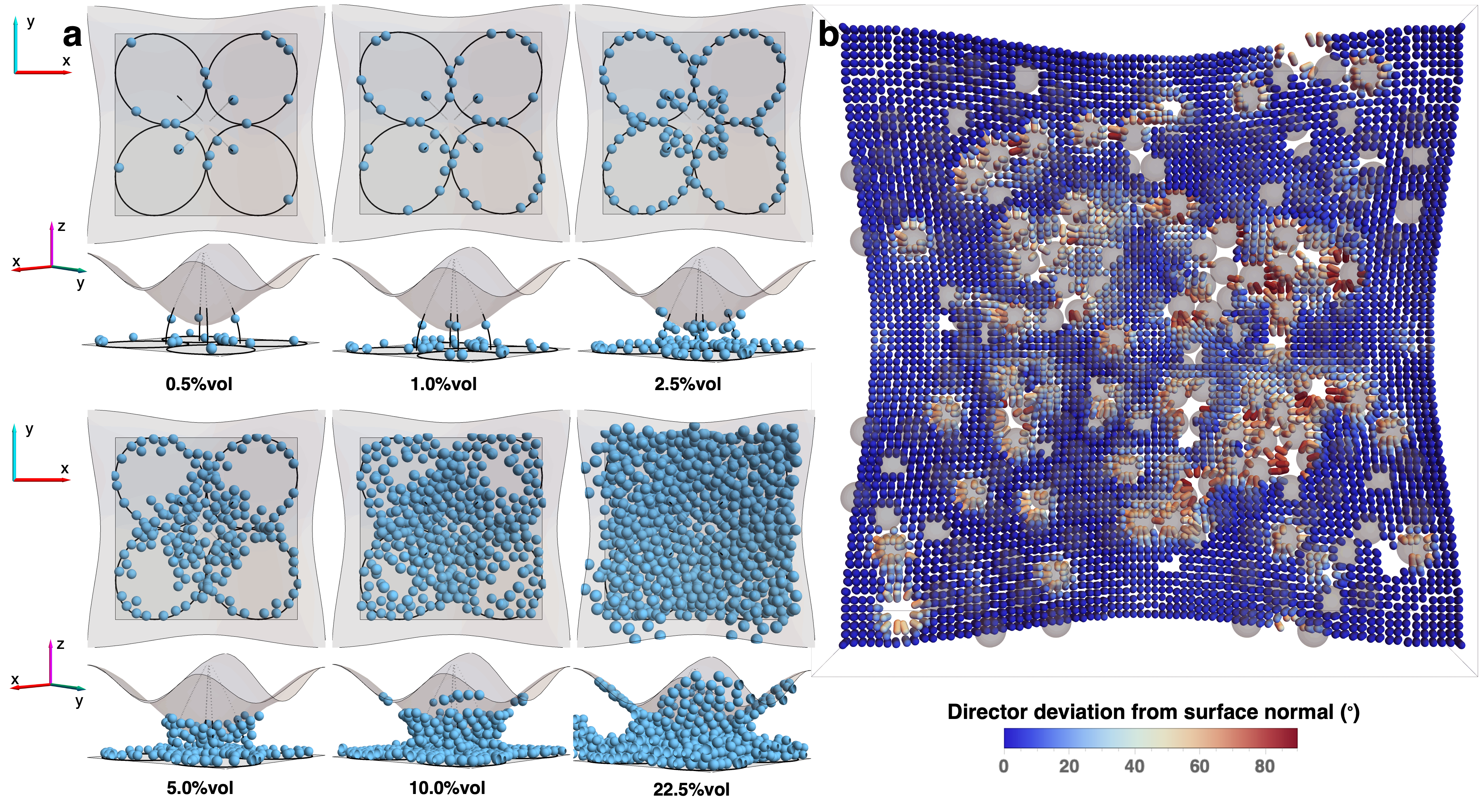}
  \caption{NP asembly and effect of NPs on anchoring. (a) Sequential assembly of NPs with increasing concentration measured in \%vol. Black lines show the positions of the elliptical and hyperbolic defects in the FCDs of the SmA phase. (b) The director field close to the undulated surface at 22.5\%vol NP concentration. Gray spheres are the NPs found close to the undulated surface.}
   \label{fig:NPAssembly}
\end{figure}

To gain insight into the concentration dependence of NP assembly in the SmA LC, we again employ LdG numerical free energy relaxation with adaptations to approximate a SmA configuration. NPs are introduced sequentially as small, spherical inclusions in the smectic-like LC configuration, with each NP's location chosen to minimize the LC free energy, with all previously inserted NPs held fixed. When planar anchoring is imposed on the NP surfaces, we observe a sequence of NP assembly behavior: the NPs first decorate the  elliptical defect lines of the FCDs, then decorate the hyperbolic defect lines, then pack near the flat surface, and next pack near the undulated surface,  before finally filling the remainder of the LC bulk. This sequence of assembly is illustrated in Figures \ref{fig:NPAssembly}-a. We also numerically investigate the effect of anchoring imposed on the surface of the NPs in the Supporting Information (Figure~S4), where we find that the ordered assembly of NPs is similarly observed in the case of homeotropic anchoring and no anchoring is imposed on the NP surface.

We can understand this assembly sequence based on the elastic energy cost of NPs, whose degenerate planar surface anchoring promotes bend distortions in the LC director field. As defects lines are regions of high free energy density,  the LC free energy favors replacing portions of defect lines with NPs or colloidal particles, essentially reducing the energetic cost of these distortion sources by overlapping them \cite{Lavrentovich2014}. Similar reasoning explains the subsequent assembly of NPs at the boundaries, as the total boundary area (including NP surfaces) is thereby reduced, representing a benefit both in surface energies and bulk elastic distortions \cite{Bitar2011,Blanc2013}.  \\

Examining the effect of NP concentration on the director field, we performed an additional LdG free energy minimization over the LC bulk, with NP locations held fixed at different volume fractions in the assembly sequence of Figure~\ref{fig:NPAssembly}-a.  At low NP concentration, the LC director near the double undulated surface primarily follows that surface's homeotropic anchoring condition. With increasing NP concentration, however, more sites near the double undulated surface are filled with NPs, resulting in an effective change in surface anchoring to the planar anchoring of the NP surfaces. This is further illustrated in Figure \ref{fig:NPAssembly}-b, where regions fully covered in NPs correspond to having $\sim 90^{\circ}$ deviation of the director from the surface normal of the undulated surface, i.e., the effective anchoring in these regions is planar. In contrast, regions of the undulated surface with smaller NP coverage correspond to regions with a small deviation of the director from the surface normal. These results support our hypothesis that high NP concentrations cause the hybrid-aligned FCD texture to be replaced with approximately vertical layers having horizontal normal direction, similar to the bookshelf arrangement of layers observed previously with F-SiO2 NPs in semi-fluorinated SmA LC \cite{Honglawan2015}.

\section{Conclusion}

In this work, we demonstrated the benefits of using 3D printing to create curved confining geometries for a SmA LC mixture in order to control its defects and manipulate NP assembly. We demonstrate how this technique offers design flexibility and scalability, enabling the development of LC films with tailored properties. We also explain how 3D printing provides a powerful tool for engineering defects in smectic LCs. Additionally, we establish how these defects could be used as a template to assemble Au NPs and discuss the features of this approach.

This technique of TSA could be adapted to various classes of functional nanomaterials useful in many technologies. For instance, it can be employed to enhance the performance of solar cells by incorporating quantum dots \cite{D0NA00169D} or nanowires \cite{cryst9020087} into self-assembled FCDs. This may improve light absorption and charge transport within solar cells, leading to enhanced energy conversion efficiency. Another example is the fabrication of high-performance light-emitting devices \cite{Yang2015}. By integrating luminescent nanomaterials into smectics, FCDs can be engineered to create ordered emission patterns. This control over defect structures could enable the production of devices with enhanced light extraction efficiency, color purity, and improved viewing angles. 

Additionally, defect assemblies in LCs offer opportunities for biosensing applications \cite{D1AN00077B}. By functionalizing the surfaces of nanomaterials with biomolecules or antibodies, FCDs can be used to capture and detect specific analytes or biological targets. The binding of these targets to the nanomaterials induces changes in the defect structures, which can be optically detected, enabling sensitive and label-free biosensing platforms. In each of these applications, the ability to control and manipulate LC defects provides a platform for the precise assembly and organization of nanomaterials, resulting in improved device performance, enhanced functionalities, and tailored properties.

\section{Experimental Section}
\subsection{3D printed double undulated surfaces}
The double undulated geometries used to confine the LC mixture are produced using the raster-scanning direct laser writing (rDLW) system built on a standard resonant-scanning two-photon microscope \cite{Pearre2019}. This technique can print objects with minimum feature sizes of about $\sim$4$\times$1$\times$1$\mu$m in the $X$, $Y$, and $Z$ directions, respectively.  The 3D printed surfaces have undulations in both $X$ and $Y$ directions with a peak-to-peak amplitude of $A$=25$\pm$2$\mu$m and a wavelength of $\lambda$=300$\pm$2$\mu$m.  The double-undulated surfaces are initially printed with an IP-Dip photoresist (from Nanoscribe,  GmbH) and are then transferred into a film of cross-linked polydimethylsiloxane (PDMS from Sigma-Aldrich).  The PDMS film is obtained by mixing the elastomer and the curing agent at a ratio of 9 parts to 1 (9:1). The mixture is then set under vacuum until almost all the bubbles disappear from the solution. Thereafter, we place the solution into the oven for about an hour at 80 $\degree$C until the PDMS becomes solid.  

\subsection{LC cell preparation}
The LC mixture studied in this work is obtained by mixing 80 wt.$\%$ of 4'-octyl-4-cyanobiphenyl (8CB,  purchased from Kingston Chemicals Limited that displays a SmA phase at room temperature) with 20 wt.$\%$ of 4-pentyl-4-cyanobiphenyl (5CB,  purchased from Kingston Chemicals Limited that displays a N phase at room temperature).  This LC mixture displays both N and SmA phases with a N-isotropic transition temperature at $T_{NI}=$39.3$\pm$0.2$\degree$C and a SmA-N transition temperature at $T_{SN}=$16.4$\pm$0.4$\degree$C. This mixture is helpful for the preparation of samples at room temperature, with and without NPs, because the N is less viscous than the SmA. The 8CB/5CB mixture is then confined between the PDMS film with the double-undulations and a coverslip treated with polyvinyl alcohol (PVA from Sigma-Aldrich).  The thickness between the peaks of the undulations and the flat coverslip is estimated to be around $h\approx$ 5$\pm$2$\mu$m.  The anchoring of the LC mixture is perpendicular in contact with the PDMS and planar at the PVA-treated coverslip.  These anchoring conditions induce a "hybrid texture" in the LC film responsible for the formation of defects in the N and SmA phases.

\subsection{Dispersion of gold (Au) nanoparticles (NPs) in the LC}
Spherical Au NPs of diameter 20$\mu$m stabilized with citrate ligands in an aqueous solution (purchased from NNCrystal) are dried under a slow airflow to eliminate water. The dry NPs are then dispersed in an ethanol solution and added to the 5CB/8CB mixture. The solvent was then evaporated under a slow airflow at 60$^{\circ}$C overnight. Finally, the samples were vacuum-dried at the same temperature for one hour to eliminate any residual solvent. The samples were used immediately after the dispersion of NPs to avoid the formation of big aggregates.  

\subsection{Characterization}
To characterize the 3D shape of the PDMS films with undulations and measure their peak-to-peak amplitude and wavelength, we used the Zeta-20 benchtop optical profiler.  The LC mixture is studied under an upright optical polarizing microscope (Leica DM6 M) in transmission mode equipped with a temperature controller (Instec TS102-mK2000A) that has a precision $\sim$ 0.1 °C. This microscope is capable of collecting and comparing data in different modes simultaneously (bright field, polarized, and dark field). It is also equipped with an automated XYZ stage useful for 3D imaging and measurements.  All images were recorded with a high-resolution digital camera (Leica DMC 5400).    

\section{Theory and Numerical Methods}
\subsection{Multi-step Landau-de Gennes free energy minimization}
The Landau-de Gennes (LdG) free energy utilizes a second-order traceless and symmetric tensorial order parameter, $Q_{ij}$, which contains information regarding the director $\hat{n}$ and the N degree of order, $S$ \cite{degennes_prost1995,kleman_lavrentovich_2003}. The $ij\text{th}$ element of the Q-tensor is given by, in the uniaxial limit, $Q_{ij} = \frac{3}{2}S\left( n_i n_j - \frac{1}{3} \delta_{ij}\right)$. In the absence of an external field, the LdG free energy has three components: phase free energy, distortion free energy, and surface free energy,
\begin{equation}
    \label{eqn:LdGfreeenergy}
    F_{\text{LdG}} = \int \mathrm{d}V \left( f_{\text{phase}} + f_{\text{dist}} \right) +\int \mathrm{d}A f_\text{surf}.
\end{equation}

In (\ref{eqn:LdGfreeenergy}), the bulk free energy density components are
\begin{equation}
    \label{eqn:phaseFE}
    f_{\text{phase}} = \frac{1}{2} AQ_{ij}Q_{ji} + \frac{1}{3}B Q_{ij}Q_{jk}Q_{ki} + \frac{1}{4} C \left( Q_{ij}Q_{ji} \right)^2,
\end{equation}
and
\begin{equation}
    \label{eqn:distFE}
    f_\text{dist} = \frac{1}{2} L_1 \frac{\partial Q_{ij}}{\partial x_k}\frac{\partial Q_{ij}}{\partial x_k} + \frac{1}{2} L_2 \frac{\partial Q_{ij}}{\partial x_j}\frac{\partial Q_{ik}}{\partial x_k} + \frac{1}{2} L_3 Q_{ij} \frac{\partial Q_{kl}}{\partial x_i}\frac{\partial Q_{kl}}{\partial x_j}.
\end{equation}
In the uniaxial limit, the distortion free energy can be written in terms of the Frank elastic constants:
\begin{equation}
    \label{eqn:frankoseen}
    f_\text{dist}^\text{FO} = \frac{1}{2} K_1 \left(\hat{n}\cdot\left(\nabla\cdot\hat{n}\right)\right)^2 + \frac{1}{2} K_2 \left(\hat{n}\cdot\left(\nabla\times\hat{n}\right)\right)^2 +\frac{1}{2} K_3 \left(\hat{n}\times\left(\nabla\times\hat{n}\right)\right)^2,
\end{equation}
where $K_1$, $K_2$ and $K_3$ are the splay, twist and bend parameters, respectively. In terms of the elastic parameters in Eq. \ref{eqn:distFE}, $L_1=2(-K_1 + 3K_2 +K_3)/27S_0^2$, $L_2 = 4(K_1-K_2)/9S_0^2$, and $L_3 = 4(K_3 - K_1)/27S_0^3$. We note that close to the SmA-N phase transition, the twist and bend parameters diverge since these two distortions are disallowed in the SmA phase \cite{kleman_lavrentovich_2003}.

For the surface free energy contribution, we consider two types of alignment: homeotropic (perpendicular) anchoring and degenerate planar anchoring. For surfaces with homeotropic anchoring, we employ a Rapini-Papoular surface free energy density \cite{Rapini1969, Nobili1992},
\begin{equation}
    \label{eqn:surfaceFEperp}
    f_\text{surf}^{\perp} = \frac{1}{2} W_0 \left(Q_{ij} - Q_{ij}^0\right)^2,
\end{equation}
where $W_0$ gives the anchoring strength, and $Q_{ij}^0$ gives the preferred surface tensorial order parameter, $Q_{ij}^0 = \frac{3}{2}S_0\left(\nu_i \nu_j  -\frac{1}{3}\delta_{ij}\right)$, where $\hat{\nu}$ is the surface normal. For degenerate planar anchoring, we follow the Fournier-Galatola formulation for the surface free energy density \cite{Fournier2005},
\begin{equation}
    \label{eqn:surfaceFEdegplanar}
    f_\text{surf}^\parallel = W_1\left(\tilde{Q}_{ij} -\tilde{Q}_{ij}^\perp\right)^2 + W_2 \left(\tilde{Q}_{ij}\tilde{Q}_{ji} - \left(\frac{3}{2}S_0\right)^2\right)^2.
\end{equation}
In the Fournier-Galatola formulation, $\tilde{Q}_{ij}= Q_{ij} + \frac{1}{2}S_0 \delta_{ij}$, and $\tilde{Q}^{\perp}_{ij}= P_{ik}\tilde{Q}_{kl}P_{lj}$, with projection operator $P_{ij}=\delta_{ij} - \nu_i\nu_j$ making $\tilde{\mathbf{Q}}^\perp$ the projection of $\tilde{\mathbf{Q}}$ onto the (local tangent) plane of the substrate.

While the LdG framework models the N phase and does not account for the presence of smectic layers, we perform numerical modeling in the N phase close to SmA-N transition as in previous works \cite{Suh2019, Gim2017}, utilizing smectic-like elasticity $K_3 > K_1$ and the N director field normal to layer configurations of  nonzero-eccentricity FCDs. 

In the first application of LdG free energy minimization, we consider a configuration of four converging FCDs in the smectic with eccentricity $e=0.5$, following the eccentricity of the FCDs observed in prior work \cite{Preusse2020}, with the ellipses parametrizing the FCDs placed on the flat surface where we impose degenerate planar anchoring. The point of convergence of the hyperbolae of the FCDs corresponds to the same $(x,y)$-value as the position of the trough of the double undulated surface, i.e., where the liquid crystal thickness is minimum. Homeotropic anchoring is imposed on the double undulated surface. To obtain a complete picture of the director field in the entire liquid crystal, we perform a N free energy minimization on the sites outside the FCDs, constrained by the fixed director field inside the FCDs.

Having obtained the N director field close to the SmA phase in the entire liquid crystal, we perform a second free energy minimization now for all the points in the liquid crystal to determine the defect evolution going into the N phase. Since we have been considering the N phase close to $T_{SN}$, we use $K_3 = 4K_1$, reflecting the divergence of the bend distortion parameter in the SmA phase. Finally, to complete the defect evolution into the N phase, we perform a third relaxation with $K_3$ closer to $K_1$ to obtain a configuration deep in the N phase.

\subsection{Numerical details}

Simulations for defect evolution were performed in a $450\times 450\times 40$ lattice, with lattice spacing equivalent to $4.5\text{nm}$ through a finite difference relaxation method \cite{Ravnik2009}. Material constants used for the phase free energy in (\ref{eqn:LdGfreeenergy}) were $A=-0.172\times 10^6 \text{J}/\text{m}^3$, $B=-2.12\times 10^6 \text{J}/\text{m}^3$ and $C=1.73\times 10^6 \text{J}/\text{m}^3$, which gives $S_0 = (-B + \sqrt{B^2-24AC})/6C \approx 0.533$. Defects are marked where the N scalar order parameter $S<0.8 S_0$. In the three-constant distortion free energy, we use the elastic constants for 8CB given in \cite{Cestari2009}, with $K_1=6\times 10^{-12} \text{N}$, $K_2 = 4\times 10^{-12}\text{N}$ and $K_3 = 4K_1$ for the first two relaxation steps. In the third relaxation step, $K_3=10\times 10^{-12} \text{N}$. For all the relaxation steps, homeotropic anchoring was imposed on the double undulated surface with anchoring strength $W_0 = 5\times 10^{-3} \text{J}/\text{m}^2$, and degenerate planar anchoring on the flat surface with anchoring strength $W_1 = 1.067\times10^{-1} \text{J}/\text{m}^3$ and $W_2=0$. These values for anchoring are consistent with the experimental ratio of anchoring extrapolation lengths to system size, $\xi_\parallel/h = 10^{-2}$ and $\xi_\perp/h = 10^{-1}$, where $h\sim 10^{-5} \mathrm{m}$. Simulations for NP assembly were performed in a $75\times 75\times 40$ lattice with the same lattice spacing as the defect evolution simulations. All material constants and surface anchoring constants are consistent with the defect evolution simulations. For elastic constants, we chose $K_3=4K_1$ for the near-smectic scenario, and the anchoring strength on the surface of the NPs is $W_\text{NP} = 2.4\times 10^{-4} \text{J}/\text{m}^2$, with the ratio of the anchoring extrapolation length and NP radius $\xi_{\text{NP}}/r_{\text{NP}} \sim 10$, consistent with the weak anchoring limit for the NP surface.

\begin{acknowledgement}

M.A.G. and A.J.R. acknowledge UMass Boston for the support and the McNair fellowship. J.B.D.M.G. and D.A.B. acknowledge the donors of the American Chemical Society Petroleum Research Fund for partial support of this research. This work was supported in part by the National Science Foundation under Grant No. DMR-2225543. Computational work was carried out at the Advanced Research Computing at Hopkins (ARCH) core facility (rockfish.jhu.edu), which is supported by the National Science Foundation (NSF) grant number OAC 1920103.

\end{acknowledgement}

\begin{suppinfo}

Supporting Information is available from the Journal Online Library.

\end{suppinfo}

\bibliography{achemso-demo}

\end{document}









\section{Single undulated surface}
\begin{figure}
    \centering
    \includegraphics[scale=0.35]{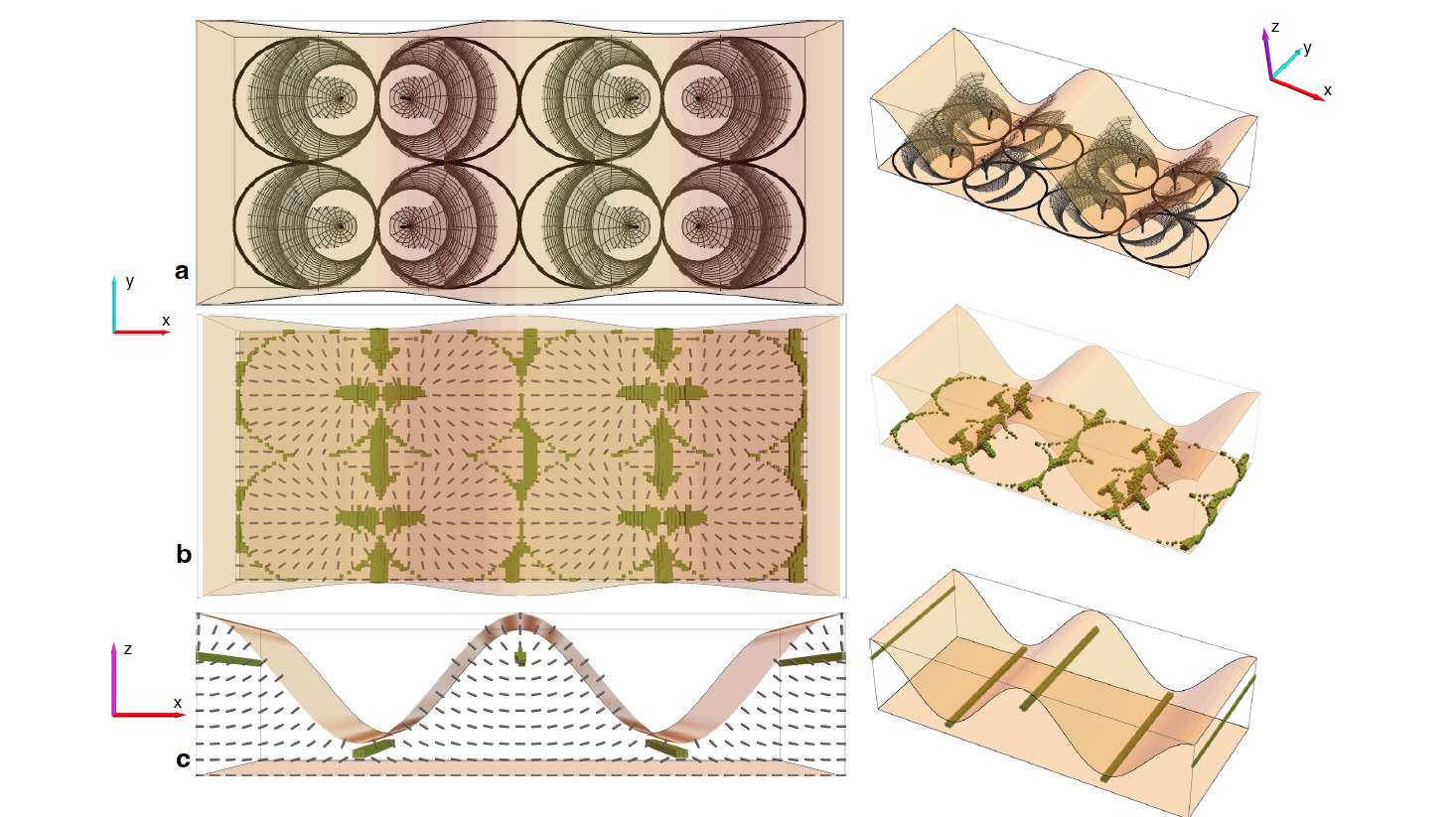}
    \caption{Numerical modeling of defect evolution in single-udnulated confinement at the smectic A-to-nematic phase transition. (a) Schematic of the positions of the FCDs used in the first application of LdG free energy minimization. (b) Defects (green) arising after first LdG free energy minimization. Gray rods represent the director field on the flat surface with degenerate planar anchoring. (c) Final defect configuration after final application of LdG free energy minimization, modeling a state in the nematic phase.}
    \label{fig:SingleUndulatedSimulation}
\end{figure}

To validate our multi-step adaptation of Landau-de Gennes (LdG) free energy minimization, and to gain more insight into how surface curvature directs defect assembly at the smectic A-to-nematic phase transition, we numerically studied the single-undulated confining geometry (oscillations along $X$ but not $Y$)  investigated experimentally in \cite{Preusse2020}. Simulations were done in a $160\times 140\times 40$ lattice with the same parameters as used in modeling the defect evolution for the double undulated surface. The system was initialized with a focal conic domain (FCD) configuration such that two sets of FCDs with eccentricity $e=0.5$ have their hyperbolic defect lines converge in pairs at $(X,Y)$ values where the thickness of the confinement geometry is minimum. Figure S\ref{fig:SingleUndulatedSimulation}-a shows the arrangement of the FCDs used as the initial condition of the model. Figure S\ref{fig:SingleUndulatedSimulation}-b shows the director configuration after first application of LdG free energy minimization. The final nematic configuration after relaxation is that of parallel wedge disclination lines running along the direction of undulation $Y$, with winding numbers alternating between $+1/2$ and $-1/2$. Disclination lines with $-1/2$ winding are found running along the troughs of the undulated surface, while disclination lines with $+1/2$ winding lie parallel to the crests of the undulated surface, as seen in Figure S\ref{fig:SingleUndulatedSimulation}-c. The observation of parallel disclination lines along the $Y$ direction is consistent with experimental findings from \cite{Preusse2020}.

\section{Formation of point-like +1 defects}
\begin{figure}
    \centering
    \includegraphics[scale=0.7]{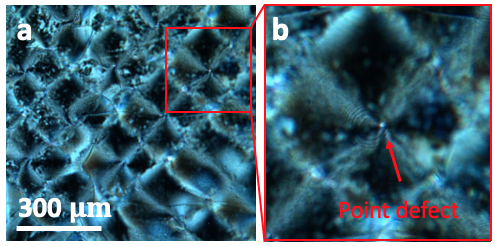}
    \caption{Formation of point-like $+1$ defects at the interstices of the ellipses of the converging FCDs, near the nematic-SmA phase transition. }
    \label{fig:pointdefect}
\end{figure}

The simulations indicate the formation of point-like $+1$ defects at the interstices of the ellipses of the converging FCDs, analogous to those seen in simulations from \cite{Suh2019}, appearing as split-core short $+1/2$ disclination lines \cite{Tasinkevych2012}. To validate our findings, we conducted experiments and analyzed the evolution of defect structures after cooling the LC mixture from the isotropic to the nematic phase, just before the phase transition to the SmA phase. The results are illustrated in Figure S\ref{fig:pointdefect}.

\section{Interaction of NPs with FCDs in flat LC films }
\begin{figure}
    \centering
    \includegraphics[scale=0.4]{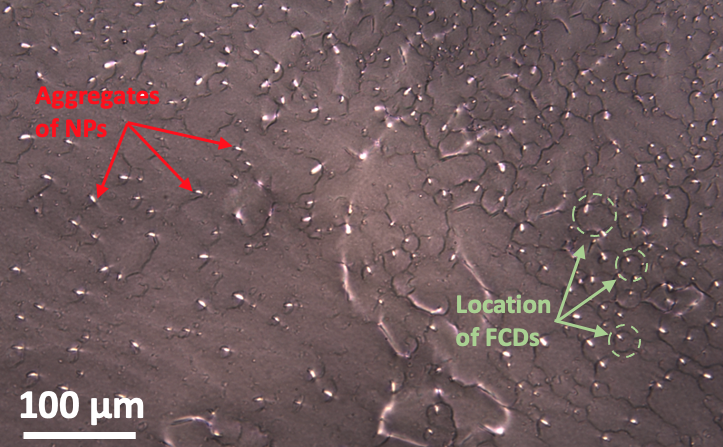}
    \caption{Aggregation of NPs in the core of FCDs. The image shows the assembly of NPs near the nematic-SmA phase transition in the same regions where the defects are created in the SmA phase. If the sample is heated to the isotropic phase, these aggregates re-disperse in the mixture. }
    \label{fig:NPsinFlatFCDs}
\end{figure}

We first analyzed the behavior of NPs in uniformly confined SmA films with hybrid alignment, where both interfaces are planar, in the presence of FCDs. This experiment aims to verify whether NPs with citrate ligands can interact with regular FCDs, before we study their interactions with FCDs at undulated surfaces. NPs were initially dispersed in the isotropic phase of the mixture and then introduced by capillarity between a PDMS film and coverslip treated with PVA to ensure hybrid texture. The sample was slowly cooled to the smectic phase until the formation of FCDs. After around 10 minutes, the mixture was heated to the nematic phase to remove the FCDs. Our results show that the NPs are accumulated in the same regions where the FCDs were created (Figure S\ref{fig:NPsinFlatFCDs}), confirming the strong interaction between the NPs and FCDs. 

\section{Anchoring on NP surface}
\begin{figure}
    \centering
    \includegraphics[scale=0.45]{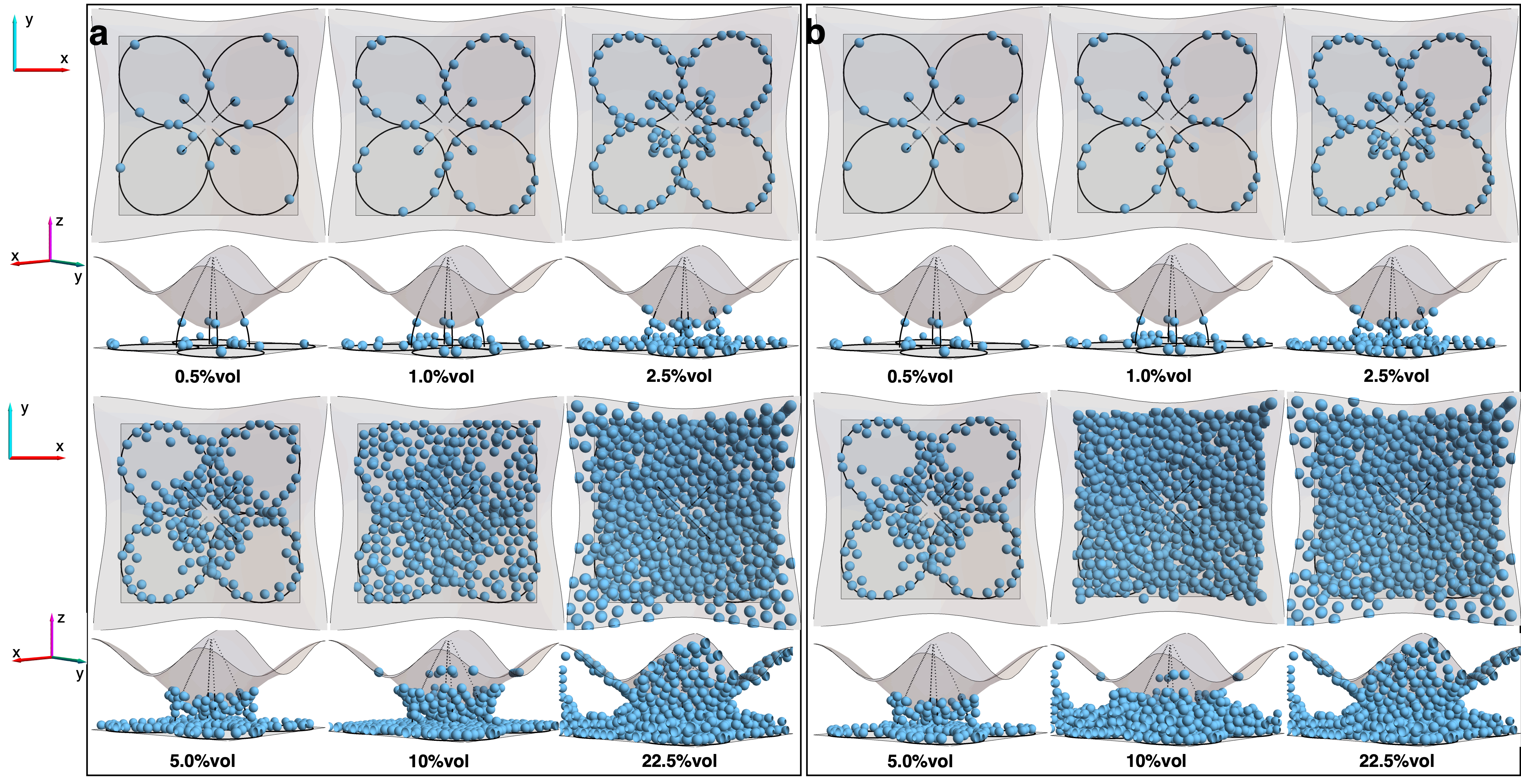}
    \caption{Sequence of NP assembly with increasing concentration for particles with homeotropic anchoring (a) and no anchoring (b) on surface.The black lines indicate the positions of the elliptical disclinations and hyperbolic defects in the FCDs of the smectic-A phase.}
    \label{fig:NPAssemblyOtherAnchoring}
\end{figure}

In addition to the degenerate planar anchoring on NP surfaces as modeled in the main text, we also numerically investigate the cases of homeotropic anchoring and no anchoring in order to demonstrate the significance of anchoring type for the assembly outcome. In simulations, a similar ordered assembly of NPs is observed when either weak homeotropic anchoring or no anchoring is imposed on the NP surfaces, as in the case of weak degenerate planar anchoring. We employ the same steps as we have done in the main text, where the anchoring strength imposed on the NP surface for degenerate planar anchoring is  $W_{\perp,\text{NP}}=2.4\times 10^{-4} \text{J}/\text{m}^2 $ (the corresponding ratio of the anchoring extrapolation length to the radius of the NP is $\xi_{\perp,\text{NP}}/r_{\text{NP}} \sim 10$). The following sequence of assembly is shown in Figure S\ref{fig:NPAssemblyOtherAnchoring}: the NPs decorate the elliptical defect lines of the FCDs, then the hyperbolic defect lines, after which the NPs pack close to the surfaces before finally filling the rest of the LC bulk.\\

\bibliography{achemso-demo}